\documentclass[11pt]{article}
\usepackage{blois,epsfig}

\def\be{\begin{equation}}
\def\ee{\end{equation}}
\def\bea{\begin{eqnarray}}
\def\eea{\end{eqnarray}}

\begin{document}
\vspace*{4cm}
\title{Holographic Hydrodynamics and Applications to RHIC and LHC}

\author{Yaron Oz}

\address{Raymond and Beverly Sackler School of Physics and Astronomy, Tel-Aviv University, Tel-Aviv 69978, Israel}

\maketitle\abstracts{
We briefly review recent developments of hydrodynamics, its gravitational description and relevance to relativistic
heavy ion collisions. We discuss the basics of hydrodynamics, the fluid/gravity correspondence, triangle anomalies and chiral effects, turbulence and
its universal structure.}

\section{Introduction}

According to the Holographic Principle \cite{'tHooft:1993gx},
the degrees of freedom of a quantum theory of gravity in a volume of space $V$ are encoded on
its boundary $A$. Thus, the Holographic Principle relates gravity in $d+1$ space-time dimensions to  field theory without gravity in one lower dimension.
In the framework of the AdS/CFT correspondence \cite{Maldacena:1997re,Aharony:1999ti}, the quantum theory of gravity is string theory on asymptotically Anti-de-Sitter (AdS) space, and the theory on the boundary is a conformal field theory (CFT).
The AdS/CFT correspondence has been generalized, and one has by now a large number of examples of non-AdS/non-CFT relations.
The non-CFT theories are QCD-like gauge field theories.
In the regime of a large field theory coupling $\lambda \gg 1$ the gravity description is weakly coupled, that is the curved geometry has a small curvature. On the other hand,
when $\lambda \ll 1$ the field theory is weakly coupled, while the gravity description is strongly coupled.

A typical curved metric that arises in this setup takes the warped form
\begin{equation}
ds^2 = d r^2 +a(r)^2(-dt^2 + dx^idx^i) \ .
\label{warped}
\end{equation}
Here $(t,x^i)$ are the field theory space-time coordinates, and the radial coordinate $r$ is interpreted as the field
theory energy (RG) scale.
The warp factor $a(r)$ encodes information about the nature of the field theory. When  $a(r)= \exp(-r)$, the metric (\ref{warped}) is that
of AdS space.
A particulary interesting regime of the field theory is the hydrodynamic approximation, where the AdS/CFT correspondence relates the field theory hydrodynamics
to perturbations of black hole (brane) gravitational backgrounds.

An important experimental framework to which the correspondence has been applied is the description of the QCD plasma produced at RHIC and LHC.
This plasma seems to exhibit a strong coupling dynamics, $\alpha_s(T_{RHIC}) \sim O(1)$.
Nonperturbative methods that can be used to study real time dynamics are largely unavailable, while
lattice QCD methods are inherently Euclidean.
The AdS/CFT correspondence provides a real-time nonperturbative framework, and one typically
uses the strong coupling properties of a CFT plasma as a
reference point for describing the strongly coupled QCD plasma.

A particulary studied quantity
is the ratio of the shear viscosity $\eta$ to the entropy density $s$, which for a generic strongly coupled gauge field theory
is low. Indeed, such a low ratio is a generic property of the gravitational description \cite{Policastro:2001yc}.

\begin{figure}[htb]
\begin{center}
\epsfig{file=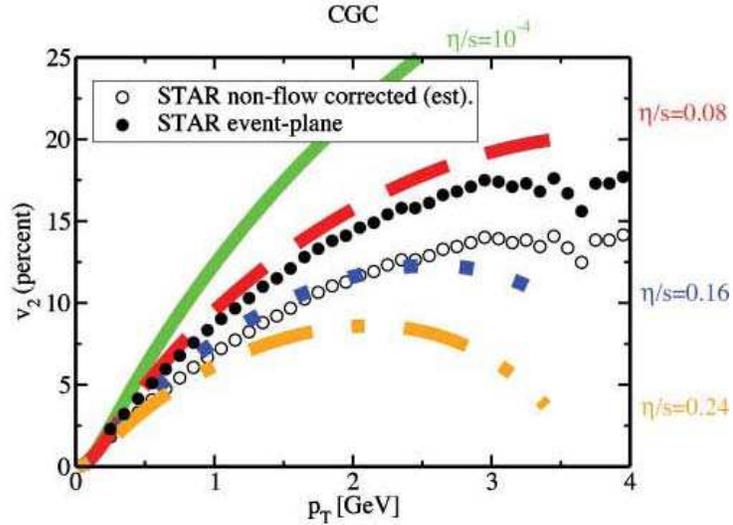,height=7cm}
\caption{Comparison between hydrodynamic simulations and experimental values of the elliptic flow parameter (Luzum,Romatschke:2008).}
\end{center}
\label{simulation}
\end{figure}

The hydrodynamic simulations at
low shear viscosity to entropy ratio are consistent with RHIC data  \cite{Luzum:2008cw}(see figure \ref{simulation}),
where the elliptic flow parameter is the second Fourier coefficient $v_2 = \langle Cos(2\phi)\rangle$ of the azimuthal momentum
distribution $dN/d\phi$
\begin{equation}
\frac{d N}{d \phi} \sim 1+  2 v_2Cos(2\phi) \ .
\end{equation}

\section{Basics of Hydrodynamics}

In the hydrodynamic regime the system is in local thermal equilibrium enforced by frequent collisions between the particles.
Thus, it is characterized by a short mean free path (correlation length), much smaller than the characteristic scale of variations of the macroscopic fields.
Since it is dominated by collisions, an appropriate description  is that of a collective fluid-type flow rather than a particle (kinetic theory) one.

\vskip 0.2cm
\noindent
{\bf The Hydrodynamic Modes:}  The effective degrees of freedom in the hydrodynamic regime are charge densities $\rho(\vec{x},t)$, which are
non-equilibrium thermal averages  $\langle  \rho(\vec{x},t)\rangle_{non-eq}$.
The hydrodynamics equations are conservation laws
\begin{equation}
\partial_{t}\rho + \partial_ij^i = 0 \ .
\end{equation}

Constitutive relations express $j^i$ in terms of $\rho$ and its derivatives. For instance, if we take
$j^i = - D \partial^i \rho$ (Fick's law), we get
\begin{equation}
\partial_t \rho  - D \partial_i\partial^i \rho = 0 \ .
\end{equation}
Writing
$\rho(\vec{k},t) = \int d^3 x e^{-i \vec{k} \cdot \vec{x}\rho(\vec{x},t)}$ we have
\begin{equation}
\rho(\vec{k},t) =  e^{ - D k^2 t} \rho(\vec{k}, t=0) \ .
\end{equation}
This is the characteristic behavior of hydrodynamic modes. It has a life $\tau(k) = \frac{1}{D k^2}$, which is infinite in the long wavelength limit $k \rightarrow 0$.
Indeed, since $\rho$ is a conserved quantity it cannot disappear locally but can only relax slowly over the entire system.

The dispersion relation for the hydrodynamic mode is $\omega = -iDk^2$. It shows up as a pole in the retarded correlation function
\begin{equation}
S(\vec{x},t) = \langle \rho(\vec{x},t),\rho(0,0) \rangle_{eq} \ ,
\end{equation}
where $\langle...\rangle_{eq}$ is thermal equilibrium average, and  we assumed  $\langle  \rho(\vec{x},t)\rangle_{eq} = 0$.
$S(\vec{x},t)$ describes the fluctuations of the charge density $\rho$ and can be measured experimentally.

\vskip 0.2cm
\noindent
{\bf Relativistic Hydrodynamic:}
We will work in four flat space-time dimensions with a Lorentzian metric $\eta_{\mu\nu} = diag(-1,1,1,1)$.
Define the hydrodynamic expansion parameter (Knudsen number) $Kn$
as $Kn\equiv l_{cor}/L \ll 1$
where $l_{cor}$ is the correlation length of the fluid, and $L$ is the characteristic scale  of variations of the macroscopic fields.
The hydrodynamics equations are conservation laws
\begin{equation}
\partial_{\mu}T^{\mu\nu} = 0,~~~~~~~~~~\partial_{\mu}J^{\mu}_a = 0 \ ,
\end{equation}
where $T^{\mu\nu}$ is the stress-energy tensor and $J^{\mu}_a$ are symmetry currents.
The constitutive relations express $T^{\mu\nu}$ and $J^{\mu}_a$ in terms of the energy
 density $\epsilon(x)$, the pressure $p(x)$, the charge densities $\rho_a(x)$ and the four-velocity field $u^{\mu}(x)=(\gamma,\gamma \beta^i)$ satisfying $u_{\mu}u^{\mu}=-1$.

Consider for simplicity neutral hydrodynamics.
The constitutive relation has the form of a series for the stress-energy tensor,
\begin{eqnarray}&&
T^{\mu\nu}(x)=\sum_{l=0}^{\infty}T^{\mu\nu}_{(l)}(x) \ , \label{series}
\end{eqnarray}
where $T^{\mu\nu}_{(l)}\sim (Kn)^l$.
The expansion is a derivative expansion, where $l$ counts the number of derivatives.
Keeping only the first term $l=0$ in the series gives ideal hydrodynamics and
the stress-energy tensor reads
\begin{equation}
T^{\mu\nu}_{(0)}= \epsilon u^{\mu}u^{\nu} + p P^{\mu\nu} \ ,
\end{equation}
where $P^{\mu\nu}= \eta^{\mu\nu}+u^{\mu}u^{\nu}$.
Since there are four conservation equations and five fields $\epsilon,p,u^{\mu}$, one more equation is needed. This is
called the equation of state $\epsilon(p)$.
For instance, in CFT hydrodynamics: $T^{\mu}_{\mu} = 0$ and the equations of state reads $\epsilon = 3 p$.
In this case, there is only one dimensionfull quantity, the temperature $T$, and $\epsilon,p \sim T^4$.
The ideal conformal fluid stress-energy tensor can be recast in the form
\begin{equation}
T^{\mu\nu}_{(0)}= T^4[\eta^{\mu\nu}+4u^{\mu}u^{\nu}] \ ,
\end{equation}
up to and overall constant coefficient.
The entropy current is defined as $S^{\mu} = s u^{\mu}$, where $s$ is the entropy density $s\sim T^3$, and
is conserved in ideal hydrodynamics $\partial_{\mu}S^{\mu} = 0$.

The first order dissipative hydrodynamics is obtained by keeping also the $l=1$ term in the
series. Going out of equilibrium, there is an ambiguity in the definition of the fields, which requires a choice of a frame.
In the case of neutral hydrodynamics there is one such frame, the Landau frame $u_{\mu}T^{\mu\nu}_{(1)} =0$. In this frame, the local rest frame of the flow is where the energy density is at rest.
The stress-energy tensor reads
\begin{equation}
T^{\mu\nu}_{(1)}= -\eta\sigma^{\mu\nu} - \zeta(\partial_{\alpha}u^{\alpha})P^{\mu\nu} \ ,
\end{equation}
where
\begin{equation}
\sigma^{\mu\nu}=\partial^{\mu} u^{\nu}+
\partial^{\nu} u^{\mu}+u^{\nu}u^{\rho}\partial_{\rho} u^{\mu}+
u^{\mu}u^{\rho}\partial_{\rho} u^{\nu}
-\frac{2}{3}\partial_{\alpha}u^{\alpha}P^{\mu\nu}
\end{equation}
is the shear tensor.
There are two transport coefficient at this order, the shear viscosity $\eta$ and the bulk viscosity $\zeta$.
Note, that in the conformal case $\zeta=0$ and the first order dissipative hydrodynamics of a CFT is determined by only one transport coefficient.
The transport coefficients can be calculated form the retarded Green functions of the microscopic thermal field theory using linear response theory
and the Kubo formula.

The entropy current is no longer conserved at the first viscous order and we have
\begin{equation}
\partial_{\mu}S^{\mu} = \frac{\eta}{2T}\sigma_{\mu\nu}\sigma^{\mu\nu} + \frac{\zeta}{T} (\partial_{\mu}u^{\mu})^2  \geq 0 \ .
\end{equation}

\section{The Fluid/Gravity Correspondence}

Consider the five-dimensional Einstein equations with a negative cosmological constant
\begin{equation}
E_{mn} \equiv R_{mn}+4g_{mn}=0,~~~~~ R=-20 \ .
\label{eqads}
\end{equation}
These equations have a  thermal equilibrium asymptotically AdS solution, the boosted black brane.
The metric in the Eddington-Finkelstein coordinates reads
\begin{equation}
g^{(0)}_{mn}dy^m dy^n  =-2u_{\mu}dx^{\mu}dr-r^2f[b r]u_{\mu}u_{\nu}dx^{\mu}dx^{\nu}+r^2P_{\mu\nu}dx^{\mu}dx^{\nu} \ ,
\label{bb}
\end{equation}
where $y=(x^{\mu}, r)$,  $u_{\mu}$ is the 4-velocity boost vector, $T=1/\pi b$ is the temperature and
$f(r)=1-\frac{1}{r^4}$.
The metric (\ref{bb}) is a solution of equations (\ref{eqads}) when $b$ and $u^{\mu}$ are constants.
It has an event horizon (null hypersurface) with planar topology at $r=b^{-1}$, whose normal is $(n^{\mu},n^r)=(u^{\mu},0)$.
The Bekenstein-Hawking entropy density $s = \frac{1}{4 b^3}$ and the normal combine to give the entropy current
$S^{\mu} =  \frac{1}{4 b^3}u^{\mu}$, where we set $G_N=1$.

The metric (\ref{bb}) is no longer a solution when $b(x^{\alpha}), u^{\mu}(x^{\alpha})$, and needs to be corrected.
One looks for a
solution of the Einstein equations (\ref{eqads}) by the method of variation of constants
\begin{equation}
g_{mn}=g^{(0)}_{mn}+ g^{(1)}_{mn} + ... \ ,
\end{equation}
where $g^{(1)}_{mn}$ includes first derivatives of $b(x^{\alpha})$ and $u^{\mu}(x^{\alpha})$.
Imposing AdS asymptotics and regularity at the horizon,
one finds that
the momentum constraint Einstein equations $E_{\mu}^r=0$ give the boundary CFT hydrodynamics equations $\partial_{\mu}T^{\mu\nu} = 0$ as a series expansion in derivatives \cite{Bhattacharyya:2008jc}.
The Hamiltonian constraint $E_r^r=0$ is the equation of state.
The ratio of the shear viscosity to the entropy density that is calculated from the gravitational background
takes the universal value $\frac{\eta}{s} = \frac{1}{4\pi}$.

The CFT hydrodynamics equations can be viewed also as the Gauss-Codazzi equations governing the dynamics of the event horizon \cite{Eling:2009sj}.
In particular, the null horizon focusing equation is equivalent to the entropy balance law of the hydrodynamic fluid \cite{Bhattacharyya:2008xc,Eling:2009sj}.

One can generalize the fluid/gravity correspondence to charged and non-conformal hydrodynamics by including bulk gauge $A_{\mu}^a$ and scalar $\phi_i$ fields, respectively.
Now the bulk viscosity is nonzero. Remarkably, one can derive a simple formula for the ratio of the bulk to shear viscosities \cite{Eling:2011ms}
\begin{equation}
\frac{\zeta}{\eta} = \sum_i \left(s \frac{d \phi^{H}_i}{ds} + \rho^a \frac{d \phi^{H}_i}{d\rho^a} \right)^2 \ ,
\end{equation}
where $s$ is the entropy density, $\rho^a$ are the charges associated with the gauge fields $A_{\mu}^a$, $\phi_i^H$ are the values of the scalar fields on the horizon,
and the derivatives are taken with couplings and mass parameters held fixed.

\section{Quantum Anomalies}

The hydrodynamic description exhibits interesting effects when a global symmetry current $J_a^\mu$ of the microscopic theory is anomalous
\begin{equation}
 \partial_\mu J_a^\mu = \frac{1}{8}C_{abc}\epsilon^{\mu\nu\rho\sigma}F^a_{\mu\nu}F^b_{\rho\sigma} \ .
\end{equation}
$C_{abc}$ is the coefficient of the triangle anomaly of the currents $J^\mu_a$,$J^\mu_b$ and $J^\mu_c$.
The form of an anomalous symmetry hydrodynamic current is modified by a term proportional to the vorticity of the fluid
\begin{equation}
\omega^\mu \equiv \frac{1}{2}\epsilon^{\mu\nu\lambda\rho}u_\nu\partial_\lambda u_\rho \ .
\end{equation}
This has been first discovered in the context of the fluid/gravity correspondence \cite{Erdmenger:2008rm}.

The global anomalous symmetry hydrodynamic current takes the form
\bea
j_a^\mu=\rho_a u^\mu +
\sigma_a{}^b\left(E_b^\mu - T P^{\mu\nu} \partial_\nu\frac{\mu_b}{T}\right) + \xi_a\omega^\mu + \xi^{(B)}_{ab} B^{b\mu} \ .
\label{anomalous}
\eea
$E^{\mu}_a = F^{\mu\nu}_au_{\nu}$, $B^{a\mu}= \frac{1}{2}\epsilon^{\mu\nu\lambda\rho}u_{\nu}F_{\lambda\rho}^a$, while $\rho_a$, $T$, $\mu_a$ and $\sigma_a^b$ are the charge densities, temperature, chemical potentials and the conductivities of the medium.
The anomaly coefficients can be calculated from the requirement that the entropy current has a positive divergence, $\partial_{\mu}S^{\mu} \geq 0$
\cite{Son:2009tf,Neiman:2010zi}.
They read
\bea
\xi_a &=& C_{abc}\mu^b\mu^c + 2\beta_a T^2 - \frac{2\rho_a}{\epsilon + p}\left(\frac{1}{3}C_{bcd}\mu^b\mu^c\mu^d + 2\beta_b\mu^b T^2 \right)\ ,  \nonumber \\
    \xi^{(B)}_{ab} &=& C_{abc}\mu^c - \frac{\rho_a}{\epsilon + p}\left(\frac{1}{2}C_{bcd}\mu^c\mu^d + \beta_b T^2\right)  \ .
\eea
The coefficients $\beta_a$ are related to gravitational anomaly \cite{Landsteiner:2011cp}, that is to the triangle anomaly diagram of the current $J^\mu_a$ and two stress-energy tensors.

Consider next possible experimental signatures of the axial current triangle diagram anomaly in a hydrodynamic description of high density QCD.
In RHIC and LHC the chemical potentials are small compared to the energy density, hence $\xi_a = C_{abc}\mu^b\mu^c, \xi^{(B)}_{ab} = C_{abc}\mu^c$.
The chiral magnetic effect and the chiral vortical effect correspond to charge separation and baryon number separation, respectively \cite{Kharzeev:2010gr}.

The chiral magnetic effect corresponds to the generation of an electric current in the direction of the magnetic field $\vec{j}\sim \mu_A\vec{B}$, using
$C_{abc}$ of two electric currents and one axial current in (\ref{anomalous}), with $\mu_A$ being the axial chemical potential.
 The chiral vortical effect corresponds to the generation of a baryon number current in the direction of the vorticity vector $\vec{j}\sim \mu_A\mu_B\vec{\omega}$, using
$C_{abc}$ of two Baryon number currents and one axial current in (\ref{anomalous}), with $\mu_B$ being  the Baryon chemical potential.

Another possible experimental signature is based on the idea that the axial charge density,
in a locally uniform flow of massless fermions,
is a measure of the alignment between the fermion spins.
When the QCD fluid freezes out and the quarks bind to form hadrons,
aligned spins result in spin-excited hadrons.
The ratio between spin-excited and low spin hadron production and its angular distribution
may therefore be used as a measurement of the axial charge distribution.
One predicts an enhancement of spin-excited hadron production along the rotation axis of the collision,
the cone compared to the belt (see figure 2) \cite{KerenZur:2010zw}.
\begin{figure}[htb]
\begin{center}
\epsfig{file=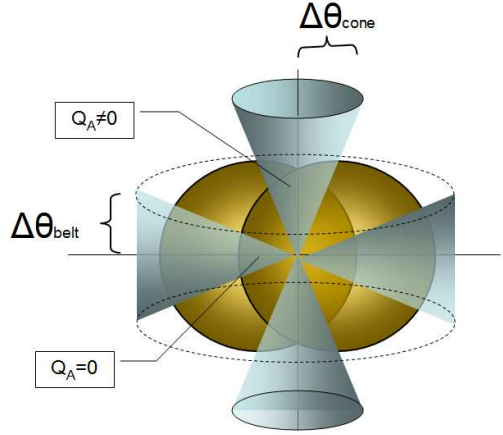,height=6cm}
\caption{An enhancement of spin-excited hadron production along the rotation axis of the collision.}
\end{center}
\label{cone}

\end{figure}

New chiral effects arise in superfluid hydrodynamics \cite{Bhattacharya:2011tr,Neiman:2011mj}, that is when there are spontaneously broken symmetries.
This is a relevant framework for QCD at high densities and low temperature \cite{Alford:2007xm}.
One possible observable effect is the chiral electric effect \cite{Neiman:2011mj}, which is a generation of an electric current perpendicular to the electric field
\begin{equation}
   J_{CEE}^{a\mu} = c^a{}_{bc}\epsilon^{\mu\nu\rho\sigma}u_\nu\xi_\rho^b E^c_\sigma \ , \label{eq:chiralE}
\end{equation}
compared to the standard electric conductivity term $J_{Conduct}^{a\mu} = \sigma^{ab}E_b^\mu$.
$c_{abc}$ are the transport coefficients proportional to the triangle anomaly diagram coefficient $C_{abc}$ ,  $\xi_\mu^a$ is the phase gradient of the broken symmetry, which is proportional to the velocity of the superfluid part.
If the broken charge is gauged as in the superconducting case, the gauge fields will be dynamically excluded from the bulk of the system. When we have unbroken gauged charges, then their associated gauge fields may enter the superfluid.


\section{Turbulence and Singularities}

The Reynolds number ${\cal R}_e$ is a dimensionless parameter that measures the relative strength of the ideal and the viscous parts
of the stress-energy tensor, and is inversely proportional to the Knudsen number.
For instance, in CFT hydrodynamics we get
\begin{equation}
{\cal R}_e \sim \frac{T L}{\eta/s} \ ,
\end{equation}
where $T$ is the temperature, $L$ is the characteristic scale of variation of the macroscopic fields and
$\eta/s$ is the ratio of the shear viscosity and the entropy density.
For a large Reynolds number ${\cal R}_e \sim 10^3$ the fluid exhibits turbulence.

Non-relativistic turbulence exhibits a universal structure as we will see in the next section.
Is there a universal structure also in relativistic turbulence?
Consider the hydrodynamics equation
with a random force term
\begin{equation}
\partial^{\nu}T_{\mu\nu}= f_{\mu} \ .
\end{equation}
In the inertial range of scales, scales much smaller than the scale of the force and much bigger than the viscous scale, one can derive
the exact scaling relation \cite{Fouxon:2009rd}
\begin{equation}
\langle T_{0j}(0, t) T_{ij}(r, t)\rangle=\epsilon r_i \ ,
\label{Tscaling}
\end{equation}
where $\langle ...\rangle$ is an average with respect to the random force and  $\epsilon$ is a nonuniversal constant.
Similarly,
in charged hydrodynamics with a conserved symmetry current $J_{\mu}$ one can derive
\begin{equation}
\langle J_{0}(0, t) J_{i}(r, t)\rangle= \epsilon r_i \ .
\label{JJ}
\end{equation}

It would be interesting to look for experimental setup where these can be verified.
For gold collisions at RHIC, the characteristic scale $L$ is the radius of a
gold nucleus $L\sim 6$ Fermi, the temperature is the QCD scale
$T\sim 200$ MeV, and $\frac{\eta}{s} \sim \frac{1}{4\pi}$ is a characteristic value of strongly coupled gauge theories.
With these $R_e$
is too small for an experimental realization of relativistic turbulence.
An experimental setup, where one may be able to study universal properties of relativistic turbulence
is condensed matter physics.
For instance, it was noticed that there is an
emergent relativistic symmetry of electrons in graphene near its quantum critical point, for which a relativistic
nearly ideal fluid description may be appropriate.

An important issue in the hydrodynamic description is whether starting with appropriate initial conditions, where the velocity vector field and its derivatives are bounded, can the system evolve such that it will exhibit within a finite time a blowup of the derivatives of the vector field.
Physically, such singularities if present, indicate a breakdown of the effective hydrodynamic description at long distances and imply
that some new degrees of freedom are required.
The issue of hydrodynamic singularities has an analogue in gravity. Given an appropriate Cauchy data, will the evolving space-time
geometry exhibit a naked singularity, i.e. a blowup of curvature invariants  and the energy density of matter fields at a point not covered by a horizon.
For a first step in relating the hydrodynamics and gravity singularities using the Penrose inequality \cite{Penrose:1973um}, see \cite{Oz:2010wz}.

\section{Nonrelativistic Flows}

Consider the equations of relativistic CFT hydrodynamics
\begin{eqnarray}
\nonumber & \partial_{\mu}u^{\mu} + 3 D ln T = \frac{1}{2\pi T}\sigma_{\mu\nu}\sigma^{\mu\nu} \ , \\
& a_{\sigma} + P_{\sigma}^{\mu}\partial_{\mu} ln T =  \frac{1}{2\pi T}P_{\sigma}^{\mu}(\partial_{\alpha}\sigma_{\mu}^{\alpha} - 3\sigma_{\mu}^{\alpha}a_{\alpha}) \ ,
\label{visceq}
\end{eqnarray}
where $D = u^{\mu}\partial_{\mu}$, $a_{\sigma}= D u_{\sigma}$ and we took $\frac{\eta}{s} = \frac{1}{4 \pi}$.

The incompressible Navier-Stokes equations can be obtained in the nonrelativistic limit of relativistic hydrodynamics \cite{Fouxon:2008tb}.
Expand $u^{\mu} = (1+v^2/2+..., v^i), T =  T_0(1+ P + ...)$ where we
scale $v^i \sim \varepsilon, \partial_i \sim \varepsilon, \partial_t \sim \varepsilon^2, P \sim  \varepsilon^2$, where $\varepsilon \sim 1/c$.

The first equation in (\ref{visceq}) gives the incompressibility condition $\partial_i v^i = 0$. The second equation
gives
\begin{equation}
\partial_t v^i + v^j \partial_j v^i = -\partial^i P + \nu \Delta v^i \ ,
\label{NS}
\end{equation}
where $\nu = \frac{1}{4 \pi T_0}$ is the kinematic viscosity.

Most nonrelativistic fluid flows in nature are turbulent. The Reynolds number takes the form
\begin{equation}
{\cal R}_e = \frac{L V}{\nu} \ ,
\end{equation}
where $L$ and $V$ are, respectively, a characteristic scale and velocity of the flow.
${\cal R}_e$ is generically large since the  kinematic viscosity is generically low. For instance,  the kinematic viscosity of water at room temperature is $\nu \simeq 10^{-6} \frac{m^2}{sec}$.

There is experimental and numerical evidence that in the inertial range of distance scales, the flows exhibit a universal behavior
\begin{equation}
S_n(r) \equiv \langle \left(({\bf v}({\bf x})-{\bf v}({\bf y}))\cdot \frac{{\bf r}}{r}\right)^n\rangle \sim r^{\xi_n} \ ,
\label{expon}
\end{equation}
where ${\bf r} \equiv {\bf x} - {\bf y}$, and the anomalous exponents $\xi_n$ are measurable real numbers.
The 1941 exact scaling result of Kolmogorov ${\xi}_3=1$ agrees well with the experimental data.
The major open problem of turbulence is to calculate the anomalous exponents $\xi_n$. It would be interesting to see if gravity can shed light on this
calculation \cite{Eling:2010vr}.

Note, that the relativistic scaling relation (\ref{Tscaling}) reduces in the nonrelativistic limit to Kolmogorov relation
\begin{equation}
\langle v_j(0, t)v_i(r, t)v_j( r, t)\rangle=\epsilon r_i \ .
\end{equation}

\section*{Acknowledgments}
The work is supported in part by the Israeli Science Foundation center of excellence, by the US-Israel Binational Science Foundation (BSF), and by the German-Israeli Foundation (GIF).

\end{document}